\documentstyle[preprint,revtex]{aps}
\begin{document}
\draft
\preprint{}
\begin{title}
 { CHIRAL SYMMETRY BREAKING\\ AND\\ PION WAVE FUNCTION}
 \end{title}
 \author{H. Mishra and S. P. Misra}
\begin{instit}{}
{Institute of Physics, Bhubaneswar-751005, India.}
\end{instit}
\begin{abstract}
We consider here chiral symmetry breaking through nontrivial vacuum
structure with quark antiquark condensates. We then relate the
condensate function
to the wave function of pion as a Goldstone mode. This simultaneously
yields the pion  also as a quark antiquark
bound state as a localised zero mode in vacuum. We illustrate the above
with Nambu Jona-Lasinio model to calculate different pionic properties
in terms of the vacuum structure for breaking of
exact or approximate chiral symmetry, as well as the condensate fluctuations
giving rise to $\sigma$ mesons.
\end{abstract}
\pacs{}
\narrowtext

\section{Introduction}
Nambu and Jona-Lasinio (NJL) had
linked chiral symmetry breaking \cite{njl} with properties of hadrons
quite sometime back with pion as the Goldstone mode \cite{mandula,bhaduri}.
However, pion is both a quark antiquark bound state and
a Goldstone mode. Hence through Goldstone theorem pion state
along with its wave function as a quark antiquark pair
should also get related to the vacuum structure.
It is surprising that this aspect is absent in the very
extensive literature on the topic \cite{njl,mandula,bhaduri,chrl}.

We consider phase transition as a vacuum realignment
with an explicit structure
and use the
techniques developed earlier \cite{hm88}. In this manner
for potential models we had obtained \cite{chrlgl} the gap equation to be
the same as derived through
Schwinger Dyson equation. Through Goldstone theorem \cite{gold} here
we had the extra feature that
the pion state as a space localised quark
antiquark zero mode of destabilised vacuum also gets determined
\cite{chrlgl,isi}.
We further discussed the effects of $approximate$
symmetry breaking where the gap
equation changes giving rise to a change in the pion wave function.
We discuss the same here for Nambu Jona-Lasinio model.
The reason
for doing the same in NJL model is its mathematical simplicity
and its present relevance in the context of
Salam Weinberg symmetry breaking and
top quark mass \cite{bardeen,rnm}.

We organise the paper as follows. In section II we consider
the vacuum structure with quark antiquark pairs using
an ansatz for the same by minimising the energy density. This
gives rise to the conventional gap equation and
involves a new description of
phase transition with an explicit construct for the destabilised
vacuum \cite {hm88}. In section III we
identify the pion  as  Goldstone mode and relate its wave function
with functions associated with the vacuum structure. In section IV we consider
the vacuum structure for NJL model when chiral symmetry is approximate.
We also derive here some familiar results of current algebra in the present
framework. In section V we calculate the pion charge radius using
the wave function determined from the vacuum structure. In section VI we
consider fluctuation of the condensate mode to give a
qualitative identification of
$\sigma$ meson. Section VII consists of discussions.

The method here consists of using equal time algebra \cite{spm78,glch,yopr85}
along with construction of the ground state through a variational
principle \cite{hm88,chrlgl,nmt}.
 \section{Chiral symmetry breaking and vacuum realignment}
 We shall now proceed in the same manner as earlier \cite{chrlgl}
 for the NJL model.
 Let us start with the effective Hamiltonian
\begin{equation}
{\cal H}(\vec x)=
{\psi (\vec x)^i}^\dagger (-i\vec \alpha \cdot {\vec \bigtriangledown}){\psi
(\vec x)}^i
+\int d\vec y {\psi^i_\alpha}^\dagger(\vec x)\psi_{\beta}^{j}(\vec x)
V_{\alpha\beta ,\gamma\delta}^{ij,kl}(\vec x -\vec y){\psi_{\gamma}^k(\vec
y)}^\dagger
\psi_\delta^l (\vec y),
\end{equation}
which has chiral invariance.
In the above $i,j$ stand for color indices and $\alpha,\beta$ stand for the
spinor
indices and { $V^{ij,kl}_{\alpha\beta,\gamma\delta} (\vec x -\vec y)$}
 is the potential.
For effective QCD based vector potential we may take
\begin{mathletters}
\begin{equation}
V^{ij,kl}_{\alpha\beta,\gamma\delta}(\vec x-\vec y)=
\delta_{\alpha\beta}\delta_{\gamma\delta}
(\frac{\lambda^a}{2})_{ij}
(\frac{\lambda^a}{2})_{kl}
V(\mid \vec x-\vec y\mid),
\end{equation}
where $\lambda^a$ are the  Gellman matrices. We may also have NJL model
when we take the contact potential as
\begin{equation}
V^{ij,kl}_{\alpha\beta,\gamma\delta}(\vec x-\vec y)=
G\left[(\gamma^0)_{\alpha\beta}(\gamma^0)_{\gamma\delta}\delta^{ij}\delta^{kl}-(\gamma^0\gamma^5)_{\alpha\beta}
(\gamma^0\gamma^5)_{\gamma\delta}(\tau^a)_{ij}(\tau^a)_{kl}\right]
\times \delta (\vec x-\vec y).
\end{equation}
\end{mathletters}
Here G is the dimensional interaction coupling constant and
$\tau^a$ 's  are the isospin matrices.

The field operators $\psi(\vec x)$ may be expanded as
\begin{equation}
\psi(\vec x)=\frac {1}{(2\pi)^{3/2}}\int \left[U_r(\vec k)c_{Ir}(\vec k)
+V_s(-\vec k)\tilde c_{Is}(-\vec k)\right]e^{i\vec k\cdot \vec x} d\vec k
\end{equation}
where $U$ and $V$ are given by
\begin{equation}
U_r(\vec k)=\frac{1}{\sqrt 2}\left( \begin{array}{c}1\\ \vec \sigma \cdot\hat k
\end{array}\right)u_{Ir} ;\quad V_s(-\vec k)=\frac{1}{\sqrt{2}}\left(
\begin{array}{c}-\vec \sigma .\hat k \\ 1 \end{array}\right)v_{Is}
\end{equation}
for $free$ chiral fields. The perturbative vacuum is defined by this basis
when we have $c_I\mid vac> = 0 $$={\tilde c_I}^{\dagger}\mid vac>$.
We next consider a trial vacuum state given as \cite{hm88,chrlgl}
\begin{mathletters}
\begin{equation}
\mid vac' > = U\mid vac>\equiv exp(B^\dagger - B)\mid vac>
\end{equation}
with
\begin{equation}
B^{\dagger}=\int {c_{Ir}(\vec k)}^\dagger {u_{Ir}}^\dagger(\vec \sigma .\hat k)
v_{Is}{\tilde c}_{Is}(-\vec k) f(\vec k) d\vec k .
\end{equation}
\end{mathletters}
 Here $f(\vec k)$ is a trial function associated as above with quark anti-quark
condensate. We may recall a similar construction in Bogoliubov Valatin
approach \cite{njl,mandula,chrl}.
We shall minimise
the energy density of $\mid vac'>$ to analyse the
possibility of phase transition \cite {hm88} from $|vac>$ to $|vac'>$.
For this purpose we first note that with the above transformation the operators
which annihilate $\mid vac'>$ are given as
\begin{equation}
b_I(\vec k)=Uc_{I}(\vec k)U^{-1},
\end{equation}
which with an explicit calculation yields the Bogoliubov transformation
\begin{equation}
\left(
\begin{array}{c} b_{Ir}(\vec k)\\{\tilde b}_{Is}(-\vec k)
\end{array}
\right)
=\left(
\begin{array}{cc}
cosf & -\frac{f}{\mid f\mid}sinf a_{rs}\\
\frac{f^*}{\mid f\mid}sinf (a^\dagger)_{sr} & cosf
\end{array}
\right)
\left(
\begin{array}{c}
c_{Ir}(\vec k)\\
{\tilde c}_{Is}(-\vec k)
\end{array}
\right ).
\end{equation}
Here $a_{rs}={u_{Ir}}^{\dagger}(\vec \sigma \cdot \hat k)v_{Is}$.
Using the above transformation (6) or (7) the expectation
value of the Hamiltonian with respect to $\mid vac'>$ is given as
\begin{equation}
{\cal E}=<vac'\mid {\cal H}(x)\mid vac'>\equiv T+V,
\end{equation}
where $T$ and $V$ are the expectation values corresponding to the kinetic
and the potential terms in Eq.(1). With a straightforward evaluation
we then obtain that
\begin{equation}
T=<vac'\mid \psi^i(\vec x)^{\dagger}(-i\alpha . {\vec
\bigtriangledown})\psi^i(\vec x)\mid vac'>
=-\frac{2N}{(2\pi)^3}\int d\vec k\mid \vec k \mid cos 2f(k),
\end{equation}
where $N=N_c\times N_f$ is the total number of quarks.
Similarly the potential term is given as
\begin{equation}
V=\frac{1}{(2\pi)^6}\int {{\tilde V}_{\alpha\beta,\gamma\delta}}^{ij,kl}(\vec
k_1-\vec k_2)
(\Lambda_{+}(\vec k_1))_{\beta\gamma}(\Lambda_-(\vec k_2))_{\delta\alpha}
d \vec k_1 d\vec k_2, \label{v}
\end{equation}
where ${\tilde V}(\vec k)$ is the Fourier transform of the potential
$V(\vec r)$ given as
\begin{equation}
{\tilde V}(\vec k)=\int V(\vec r)e^{i\vec k . \vec r}d \vec k,
\end{equation}
and $\Lambda_{\pm}$ are
\begin{equation}
\Lambda_{\pm}(\vec k)=\frac{1}{2}\left(1\pm \gamma^0 sin2f(k)\pm
\vec \alpha .\hat k cos2f(k)\right).
\end{equation}
The expression for $V$ as in Eq.(~\ref {v}) can be calculated for Eq.(1) or
(2a) for an effective potential \cite{mandula}. We shall however
now illustrate the method with Eq.(2b) for NJL model corresponding
to the contact potential.
The total energy density then becomes
\begin{equation}
{\cal E}\{f\}={\cal E}=-\frac {2N}{(2\pi)^3}\int d\vec k
\mid\vec k \mid cos2f - 2GN(2N+1)I^2
\end{equation}
with
\begin{equation}
I=\frac {1}{(2\pi)^3}\int sin 2f(k) d\vec k. \label {i}
\end{equation}
The leading order in N here corresponds to the Hartree approximation
\cite{chrl}.
The energy functional ${\cal E}\{f\}$ here is quadratic in $f(k)$ and it is
to be
determined by minimising the energy density.
This yields that
\begin{equation}
tan 2f(k)=\frac {2GI(2N+1)}{k}\equiv M/k,
\end{equation}
where, $M \equiv 2GI(2N+1)$ is the dynamically generated mass.
Further, substituting the above in Eq.(~\ref{i}) yields the self
consistency relation
\begin{equation}
M=\frac {2G(2N+1)}{(2\pi)^3}\int ^{\Lambda} \frac {M}{\sqrt{M^2+k^2}}
d \vec k,\label {m}
\end{equation}
with $\Lambda$ above as the ultraviolet cutoff for NJL model. Eq.(\ref{m})
is usually derived through an approximate solution
to Schwinger Dyson equation \cite
{bhaduri}. We followed here an alternative variational method with
phase transition as {\it { vacuum realignment}}
 as in Eq.(5), determined through
 {\it{ minimising energy density functional}} \cite{hm88}.

The above equation has a solution with $ M \not = 0$ (Goldstone phase)
provided
\begin{equation}
G\Lambda^2(2N+1)>2\pi^2. \label{G}
\end{equation}
The energy density of $\mid vac'>$ with respect to the perturbative vacuum
$\mid vac>$ may be evaluated to be
\begin{equation}
\Delta {\cal E} = {\cal E}\{f\}-{\cal E}\{f=0\}
=\frac{2N}{(2\pi)^3}\int^{\Lambda} (k-\sqrt{k^2+M^2})d\vec k +
\frac{N}{2G(2N+1)}M^2
\end{equation}
 which is negative when a nontrivial solution to
Eq.(~\ref {m}) exists or Eq.(~\ref{G}) is satisfied. The state with condensates
$\mid vac'>$  then becomes the physical vacuum.
One may also calculate the  order parameter $<\bar \psi \psi>$ given as
\begin{equation}
<vac'\mid \bar \psi \psi\mid vac'>=-\frac{1}{(2\pi)^3}\times 2NM
\int^ \Lambda \frac{d\vec k }{\sqrt{k^2+M^2}}
\end{equation}

\section {Goldstone theorem and pion wave function}

We shall now recapitulate \cite{chrlgl} that the present
description of phase transition
permits us to define pion also as a quark antiquark pair.
 From the gap equation we obtained two solutions for the field operators
corresponding to $sin2f(k)=0$ or $sin2f(k)\not = 0$ along with the
corresponding ground state as $\mid vac>$ or $\mid vac' >$ respectively.

For the case of chiral symmetry breaking,
we have the gap equation
\begin{equation}
1=\frac {2G(2N+1)}{(2\pi)^3}\int ^{\Lambda} \frac {1}{\sqrt{M^2+k^2}}
d \vec k,\label {mm}
\end{equation}
which determines the value of the mass parameter $M$. Once $M$ is
determined, the function $f(\vec k)$ becomes known and hence the
condensate structure of vacuum becomes known. However, the Hamiltonian
of equation (1) had chiral symmetry, which through equation (19)
or otherwise is now seen to be broken. Hence we should have a
Goldstone mode corresponding to a zero mass particle \cite{gold}.
We shall approach this theorem in a modified manner to obtain the
wave function as a quark antiquark pair \cite{isi}.
When chiral symmetry remains good,
\begin{equation} {Q_5}^a \mid vac >=0 \end{equation}
where ${Q_5}^a$ is the chiral charge operator given as
\begin{equation}
{Q_5}^a =\int \psi(\vec x)^{\dagger}\frac{\tau^a}{2}\gamma ^5\psi(\vec x)
d \vec x.
\end{equation}
For symmetry broken case however
\begin{equation}
{Q_5}^a\mid vac'>\not = 0.
\end{equation}
We expect that this will describe a pion of zero total momentum.
Since it will be massless it will also have zero energy.
corresponding to the pion state. To show this we first note that
\begin{equation} \left[{Q_5}^a, H\right]=0
\end{equation}
irrespective of whether ${Q_5}^a$ and $H$ are written in terms of
field operators corresponding to $sin 2f=0$ or $sin 2f\not =0$
since the anticommutation relation between the operators remain unchanged
by the Bogoliubov transformation.
Clearly, for the Goldstone phase, $|vac'>$ is an approximate
eigenstate of $H$ with
${\cal E} V$ as the approximate eigenvalue (V being the total volume).
With $H_{eff}=H-{\cal E} V$, we then obtain from Eq.(23) that
\begin{equation}
H_{eff}{Q_5}^a|vac'>=0
\end{equation}
i.e. the state ${Q_5}^a\mid vac'>$ with zero momentum
has also zero energy, thus corresponding to the massless pion.
Explicitly, using Eq.(3) and Eq.(7), we then obtain
with $q_I$ now  as two component isospin doublet corresponding to
 (u,d) quarks above,
\begin{equation}
\mid \pi^a(\vec 0)>=N_{\pi}\int q_I(\vec k)^\dagger (\frac{\tau^a}{2})
{\tilde q_I}(-\vec k)sin 2f(k) d\vec k \mid vac'>, \label{pi0}
\end{equation}
where, $N_\pi$ is a normalisation constant.
The wave function for pion thus is given as proportional to
$\tilde u(\vec k)\equiv sin2f(k)$.
The isospin and spin indices of $q^\dagger$ and $\tilde q$ for
quarks have been suppressed. Further, with
\begin{equation}
<\pi^a(\vec 0)\mid \pi^b(\vec p)>=\delta^{ab}\delta(\vec p),
\end{equation}
the normalisation constant $N_\pi$  is given by
\begin{equation}
{N_{\pi}}^2 \times \frac{N_cN_f}{2}\int sin^2 2f(k) d\vec k =1.
\end{equation}
Clearly the state as in Eq.(~\ref {pi0}) as the Goldstone mode
will be
accurate to the extent we determine the vacuum structure sufficiently
accurately through
variational or any other method so that $|vac'>$ is
 an eigenstate of the Hamiltonian. The above results relate pion
wave function with vacuum structure for any example of chiral
symmetry breaking.

We may note that we could relate pion wave function to the vacuum
structure since vacuum had an explicit structure as in equations (5).
In fact, the two body condensate as in equations (5) for the
destabilised  vacuum is strictly related to the presence of the
zero mode as seen here.

\section{Approximate chiral symmetry}

While considering chiral symmetry breaking, we often use results from
current algebra so that we may obtain numbers for approximate chiral
symmetry breaking. For the sake of completeness, with the present
mechanism, we elaborate \cite{chrlgl} these results so as to use
the same for NJL model. For this purpose we may add
a small mass term to the
Hamiltonian that breaks the chiral symmetry explicitly. Then ${Q_5}^a
\mid vac'>$ will not be a zero mode and will have finite mass. In fact
the mass of the pion in the lowest order will now be $m_\pi$ formally given as
\begin{equation}
<\pi^a(\vec 0)\mid H_{sb}\mid \pi^a(\vec 0)>=m_{\pi}\delta(\vec 0),
\label{mpi}
\end{equation}
where $H_{sb}$ is the symmetry breaking part of the Hamiltonian
corresponding to the Hamiltonian density
${\cal H}_{sb}=m{\bar\psi}\psi$,  $m$ being the current quark mass.
The above may be related to $N_\pi$ and pion decay constant as
 follows. Firstly
we note that the identity for pion decay constant is \cite{sakurai}
\begin{equation}
<0\mid {{J_5}^0}^a\mid\pi(\vec p)>=\frac{i}{(2\pi)^{3/2}}
\times\frac{c_\pi\times p_0}{\sqrt{2p_0}}\times
e^{i\vec p.\vec x}, \label{sak}
\end{equation}
where, $c_\pi=94 MeV$. The normalisation constant
$N_\pi$ in Eq.(\ref{pi0}) is then given by using
\begin{eqnarray}
{N_\pi}^{-2}\times \delta(\vec 0) & = &<vac'\mid {Q^a}_5
{Q^a}_5\mid vac'>\nonumber\\ & = &
\int <vac'\mid {Q^a}_5\mid \pi^b(\vec p)>d\vec p<\pi^b(\vec p)\mid
{Q^a}_5\mid vac'>,
\end{eqnarray}
where we have saturated the intermediate states with pions.
The index $b$ is summed and there is no summation over the index $a$.
With Eq.(\ref{sak}) and Eq.(31)
 we then have
\begin{equation}
{N_{\pi}}^{-2}=\frac {1}{2}\cdot(2\pi)^3\cdot{m_\pi{c_{\pi}}^2},
\end{equation}
which links $N_{\pi}$ of vacuum structure with pion mass
and pion decay constant.
We shall now substitute explicitly the pion state as in equation (26)
in equation (29). We shall also substitute the value of
normalisation constant in equation (26) by equation (32). On using
straightforward commutation relations, we then obtain that
\begin{eqnarray}
& & <\pi^a(\vec 0)\mid H_{sb}\mid\pi^a(\vec 0)> =
\frac{2}{m_\pi {c_\pi}^2}\times\frac{1}{(2\pi)^3}\cdot
<vac'\mid{Q^a}_5 H_{sb} {Q^a}_5\mid vac'>
\nonumber\\& = &
\frac{2}{m_\pi
{c_\pi}^2}\times\frac{1}{(2\pi)^3}\times\frac{1}{2}<vac'\mid\left[\left[{Q^a}_5,H_{sb}
\right], {Q^a}_5\right]\mid vac'>\\ & = &
\frac{2}{m_\pi {c_\pi}^2}\times -\frac{m}{2}<vac'|\bar \psi\psi\mid vac'>
\times \delta(\vec 0).
\end{eqnarray}
 From equation (34) and equation (29) we then obtain that
\begin{equation}
{m_\pi}^2=-\frac {m}{{c_\pi}^2}<\bar \psi\psi>.\label{calg}
\end{equation}
which is the familiar result for  current algebra.

\section{Charge radius of pion}

With the wave function of the pion as above, we may next estimate the size of
the Goldstone pion as related to
the vacuum structure. The pion state
with momentum $\vec p$ using translational invariance from Eq.(26) becomes
\begin{equation}
|\pi^+(\vec p)>= {N_\pi}\int d\vec k {q_I}^i(\vec k+
\frac{\vec p}{2})^\dagger (\tau^+)_{ij}{\tilde q_I}^j
(-\vec k+\frac{\vec p}{2})
\tilde u(\vec k) |vac'>.
\end{equation}
In Breit frame the electric form factor is given by \cite{spm78}
\begin{equation}
G_E(t)=(2\pi)^3<\pi^+(-\vec p)|J_0|\pi^+(\vec p)>
\end{equation}
where $t=-4p^2$ and $J_0=e\psi^\dagger\psi$.
This may be evaluated directly as
\begin{eqnarray}
G_E(t)&=& e {N_\pi}^2\times
\int d\vec k \tilde u(\vec k-\frac{\vec p
}{2})^*\tilde u(\vec k +\frac{\vec p}{2})\nonumber \\
& \times & \left\{u_1(\vec k-\vec p)u_1(\vec k
+\vec p)+(k^2-p^2)u_2(\vec k -\vec p)u_2(\vec k+\vec p)\right\}.
\label{get}
\end{eqnarray}
In the above
\begin{equation}
u_1(\vec k)=\sqrt{\frac{(1+sin2f
(|\vec k|))}{2}} ;\qquad u_2(\vec k)=\frac{1}{ |\vec k|}\sqrt{\frac{
(1-sin2f(|\vec k|))}{2}}.
\end{equation}
To calculate the charge radius we expand the above in powers of $\vec p$
and the coefficient of $p^2$ will be related to the charge radius through
\begin{equation}
G_E(t)=e(1+\frac{1}{6}R_{ch}^2 t +\cdots).
\end{equation}
With $G_E(t)$ as in Eq(\ref{get}) we then obtain that
\begin{eqnarray}
<R_{ch}^2>& = & \frac{1}{2}\int d\vec k \bigg[\frac{1}{4}({u_0'(k)}^2-
\frac{2}{k}u_0'u_0-u_0''u_0)\nonumber\\
&  & +u_0^2\left\{(u_1'^2-\frac{2}{k}u_1'u_1-u_1''u_1)+
3u_2^2+k^2(u_2'^2-\frac {2}{k}u_2'u_2-u_2''u_2)\right\}\bigg],
\end{eqnarray}
where we have substituted
\begin{equation}
u_0(k)=\frac{1}{\sqrt{(\int sin^2 2f(k) d\vec k)}}\times \tilde u(k),
\end{equation}
and, primes denote differentiation with respect to $k$.

The above formula applies for any known vacuum
realignment with condensates.
Let us now estimate the charge
 radius in Nambu Jona-Lasinio model.
We shall also use Eq.(34) for pion decay constant sothat chiral symmetry
was approximately true.
With ${\cal H}_{sb}= m\bar\psi\psi$, the extra contribution
to the energy density is $-m\times 2NI$. On
extremisation the modified gap function is given by
\begin{equation}
tan2f(k)=\frac{2G(2N+1)I+m}{|k|}=\frac{M'}{k},\label{mp}
\end{equation}
where $M'=2G(2N+1)I+m$ satisfies the equation parallel to Eq.(\ref{m})
given as
\begin{equation}
1=\frac{2G(2N+1)}{(2\pi)^3}\times\int^\Lambda \frac{d\vec k}{\sqrt{k^2+M'^2}}
+m/M'.
\end{equation}
Thus here we  may also have a vacuum realignment.

We shall now choose an optimal set of parameters for $G\Lambda^2$, $\Lambda$
and
$m$. Then, for example, with
$\Lambda=420$ MeV, $G\Lambda^2$=2.24 and m=16 MeV, we get
 $M$=305 MeV, $<-\bar\psi\psi>^{1/3}$= 220 MeV,
$R_{\pi}^2$=0.25 fermi$^2$ and $c_\pi$= 94 MeV.
Here we have taken $m_\pi$ = 138 MeV.
As a further illustration to see how the the corresponding quantities
change with parameters, for
 $\Lambda$=500 MeV, $G\Lambda^2$=2.15, $m$=10 MeV, we have,
$M$=320 MeV, $<-\bar\psi\psi>^{1/3}$=255 MeV, $R_\pi^2$=0.20 fermi$^2$ and
$c_\pi$=93 MeV. We note that the pion structure as arising from
vacuum realignment appears to give a smaller value of
charge radius than would be expected. In fact, with
$G\Lambda^2$ = 2.0,
$\Lambda$ = 700 MeV  and $m$= 5 MeV \cite {bhaduri},
we obtain that $M$= 360 MeV, $<-\bar\psi\psi>^{1/3}$= 342 MeV, $R_{\pi}^2$=
0.13 fermi$^2$ and $c_{\pi}$=103 MeV.
Thus the above set of parameters do not appear to be
acceptable \cite{com}. We may also note that the four component
Dirac field operators for the quarks will change the above numbers
as examined elsewhere \cite{glch}, which however does not
change the above remarks.
The above illustrates the nature of constraints derived for symmetry breaking
through determination of pion wave function.
A parallel approach with Bogoliubov transformations and Schwinger Dyson
equation has been used to obtain Salpeter wave function for the pion, which,
however
does not permit the definition of pion as a state since the wave function
is not normalisable and therefore did not give rise to such constraints.

\section{New modes in vacuum}
When vacuum has a structure, there can be excitations present due to
such a structure. For chiral symmetry breaking, let us substitute
\begin{eqnarray}
\bar\psi (\vec x)\psi (\vec x)& = & <\bar\psi (\vec x)\psi (\vec x)>+
M_{sc}^2\sigma(\vec x)\nonumber\\ & \equiv & \mu^3+M_{sc}^2\sigma (\vec x),
\end{eqnarray}
where $M_{sc}$ is a mass parameter and $\sigma (\vec x)$ may
represent the scalar field of vacuum fluctuations.
Then $\sigma (\vec x)$ can represent quantum fluctuations of the condensate.
In fact, we may evaluate
\begin{eqnarray}
&&<vac'|(\bar \psi (\vec x)\psi (\vec x)-\mu^3)
(\bar \psi (\vec y)\psi (\vec y)-\mu^3)|vac'>\nonumber\\
&=&  M_{sc}^4 <vac'|\sigma(\vec x)\sigma(\vec y)|vac'>\nonumber\\
&\simeq & \frac{M_{sc}^4}{(2\pi)^3}\int
\frac{e^{i\vec k .(\vec x-\vec y)}}
{2({\vec k}^2+m_\sigma ^2)^{1/2}} d\vec k,
\label{sigma1}
\end{eqnarray}
where we approximate $\sigma(\vec x)$ by a free field of
mass $m_\sigma$. Let us define
\begin{equation}
I(\vec k)= \int d\vec x \exp (-i\vec k . \vec x)
<vac'|(\bar \psi (\vec x)\psi (\vec x)-\mu^3)
(\bar \psi (\vec 0)\psi (\vec 0)-\mu^3)|vac'>
\label{sigma2}
\end{equation}
In that case, clearly free field approximation for $\sigma(\vec x)$
corresponds to
\begin{equation}
I(\vec k)=\frac{M_{sc}^4}{2\sqrt{m_\sigma ^2+\vec k^2}}.
\label{sigma3}
\end{equation}
Explicit evaluation of the left
hand side of equation (\ref{sigma2}) in the limit of small
$|\vec k|$ yields
\begin{eqnarray}
I(\vec k)&\simeq& \frac{1}{\pi^2}\bigg [ \bigg(\frac{\Lambda^3}{3}
-M^2\Lambda+M^3 tan^{-1}\big(\frac{\Lambda}{M}\big)\bigg )\nonumber\\
&-& {\vec k}^2\bigg \{\frac{\Lambda^5}{6(\Lambda^2+M^2)^2}
+\frac{1}{8}.\frac{\Lambda^3 M^2}{(\Lambda^2+M^2)^2}
+\frac {3}{16}.\frac {\Lambda M^2}{(\Lambda^2+M^2)}
-\frac {3}{16} . M tan^{-1}(\frac{\Lambda}{M})\bigg\}
\bigg ] \nonumber\\
&=&\frac{1}{8\pi^3}\bigg[\frac {M_{sc}^4}{2m_\sigma}
-{\vec k}^2.\frac {M_{sc}^4}{4m_{\sigma}^3}\bigg],
\label{sig}
\end{eqnarray}
where, in equation (\ref{sigma3}) we have kept terms upto ${\vec k}^2$.
Equating equal powers of $\vec k$ in equation (\ref{sig}) and eliminating
$M_{sc}$ in favor of $m_\sigma$ yields, with $x=M/\Lambda$,
\begin{equation}
m_\sigma^2=\frac{\Lambda^2}{2}\times
\bigg [ \frac {\frac{x^3}{3}-x^2+x^3 tan^{-1}
(\frac{1}{x})}{\frac{1}{6}\frac{1}{(1+x^2)^2}-
\frac{3}{16}x tan^{-1}(\frac{1}{x})+\frac{3}{16}\frac{x^2}{(1+x^2)}
+\frac{1}{8}\frac {x^2}{(1+x^2)^2}}\bigg ].
\end{equation}
We may next estimate the
mass of such a mode for different values of $\Lambda$ and $M$
obtained in the previous section. For example,
for $\Lambda=420$ MeV and $M=305$ MeV, $m_\sigma\simeq 2.07 M$;
for $\Lambda=500$ MeV and $M=320$ MeV, $m_\sigma\simeq 2.27 M$ and
for $\Lambda=700$ MeV and $M=360$ MeV, $m_\sigma\simeq 2.65 M$. These
may be compared with the mass of $\sigma$ field obtained through
an approximate determination of the pole
of the propagator with polarisation insertion, which is given as
$m_\sigma=(4M^2+m_\pi^2)^{1/2}$ \cite{chrl}.

\section{Discussions}
We thus consider here chiral symmetry breaking
as a vacuum realignment with an explicit construct for destabilised vacuum.
The new feature of this approach \cite{hm88} is that it enables us to
relate the function that describes
the vacuum structure determined variationally to the wave function of the
pion as the localised Goldstone mode in a straightforward manner.
This language is not only physically appealing
reproducing the conventional results but also
puts severe constraints
on the parameters for symmetry breaking as
illustrated here for NJL model. Some other aspects of low energy
hadronic properties as related to the vacuum structure for chiral
symmetry breaking have been discussed elsewhere \cite{glch}.

\acknowledgements
The authors are thankful to Snigdha Mishra, Amruta Mishra, S. N. Nayak and
P. K. Panda for many useful discussions. SPM acknowledges to Department of
Science and Technology, Government of India for the research grant
SP/S2/K-45/89 for financial assistance.


\begin{references}
\bibitem{com} There is a disagreement over the value of $\Lambda$.
Further, Eq.(17) and the corresponding equation by Bernard in Ref.3
have a difference of a factor 2. We presume that this is a printing
error since $G\Lambda^2$ they have taken does not satisfy their bound
where as it is consistent with Eq.(17).
\end{references}
\end{document}